\documentclass[aps,twocolumn,groupedaddress,floatfix]{revtex4-1}
\usepackage{amssymb,amsmath}
\usepackage{graphicx}
\usepackage{subfigure}
\usepackage[english]{babel} 
\usepackage{float}
\usepackage{color}

\usepackage{lipsum}
\begin{document}
\newcommand{\be}{\begin{equation}}
\newcommand{\ee}{\end{equation}}
\newcommand{\rojo}[1]{\textcolor{red}{#1}}

\title{Fractional dynamics in nonlinear magnetic metamaterials}

\author{Mario I. Molina}
\affiliation{Departamento de F\'{\i}sica, Facultad de Ciencias, Universidad de Chile, Casilla 653, Santiago, Chile}

\date{\today }

\begin{abstract} 
We examine the existence of nonlinear modes and their temporal dynamics, in arrays of split-ring resonators, using a fractional extension of the Laplacian in the evolution equation. We find a closed-form expression for the dispersion relation as a function of the fractional exponent as well as an exact expression for the critical coupling between rings, beyond which no fractional magnetoinductive wave can exist. We also find the low-lying families of bulk and surface nonlinear modes and their bifurcation diagrams. Here the phenomenology is similar for all exponents and resembles what has been observed in other discrete evolution equations, such as the DNLS. 
The propagation of an initially localized magnetic excitation is always ballistic, with a `speed' that is computed in exact form as a function of the fractional exponent. For a given exponent, it increases with an increase in coupling up to a critical coupling value, beyond which the ballistic speed could diverge inside the fractional interval $[0,1]$. Examination of the  modulational instability shows that it tends to increase  with an increase in the fractional exponent, where the decay proceeds via the formation of  filamentary structures that merge eventually and form pure radiation. The dynamical selftrapping around an initially localized excitation increases with the fractional exponent, but it also shows a degree of trapping in the linear limit. This trapping increases with a decrease in the exponent and can be explained by near-degeneracy considerations.

\end{abstract}

\maketitle

{\em Introduction}. Metamaterials constitute a class of man-made materials that are characterized by having enhanced thermal, optical, and transport properties that make them attractive candidates for current and future technologies. Among them, we have magnetic metamaterials (MMs) that consist of artificial structures whose magnetic response can be tailored to a certain extent. A simple realization of such a system consists of an array of metallic split-ring resonators (SRRs) coupled inductively\cite{SRR1, SRR2, SRR3}. This type of system can, for instance, feature negative magnetic response in some frequency window, making them attractive for use as a constituent in negative refraction index materials\cite{negative_refraction}. In order for SRRs to be practical, one must overcome the problem of ohmmic and radiative loss. A possible solution that has been considered is to endow the SRRs with external gain, such as tunnel (Esaki) diodes\cite{losses1,losses2} to compensate for such losses. The theoretical treatment of such structures relies mainly on the effective-medium approximation where the composite is treated as a homogeneous and isotropic medium, characterized by effective macroscopic parameters. The approach is valid, as long as the wavelength of the electromagnetic field is much larger than the linear dimensions of the MM constituents. The simplest model uses an array of split-ring resonators (Fig.1), with each resonator consisting of a small, conducting ring with a slit. Each SRR unit in the array can be mapped to a resistor-inductor-capacitor (RLC) circuit featuring self-inductance $L$, ohmic resistance $R$, and capacitance C built across the slit. In our case, we will consider the case of negligible resistance $R$, and thus each unit will possess a resonant frequency $\omega_{0}\sim 1/\sqrt{L C}$.

In the absence of driving and dissipation, the dimensionless evolution equations for the charge $q_{n}$ residing at the $nth$ ring are
\be
{d^2\over{d t^2}}\left( q_{n} + \lambda (q_{n+1} + q_{n-1})\right) + q_{n} + \chi\ q_{n}^3 = 0 \label{eq1}
\ee
where $q_{n}$ is the dimensionless charge of the nth ring,  $\lambda$ is the coupling between neighboring rings which originates from the dipole-dipole interaction, and we have included nonlinear effects, whose strength is controlled with the nonlinear parameter $\chi$. This effect is present when the space in the slit of each ring, is filled with a nonlinear (Kerr) dielectric.
\begin{figure}[t]
 \includegraphics[scale=0.14]{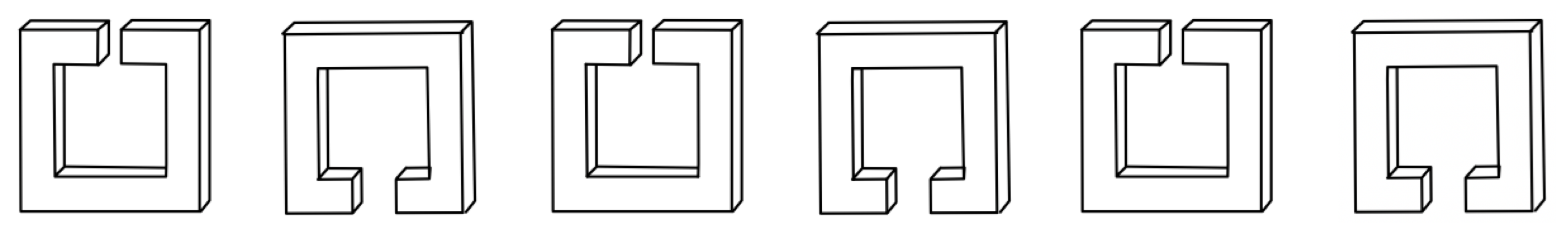}\\
  \includegraphics[scale=0.14]{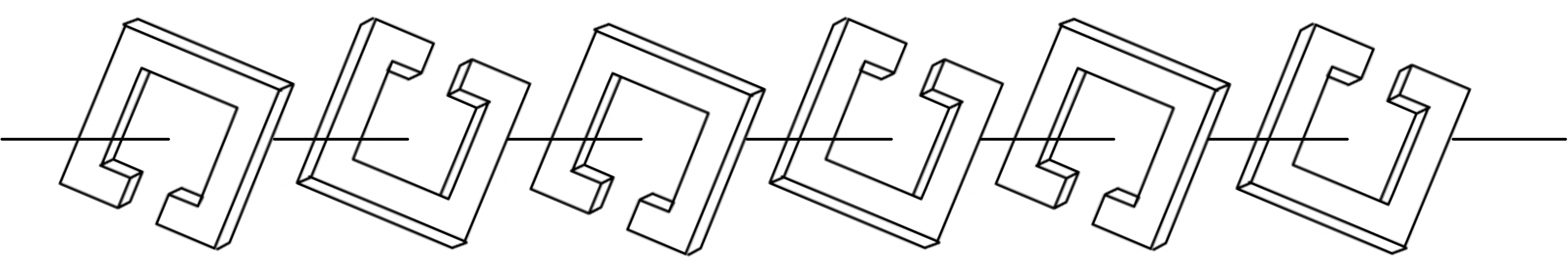}
  \caption{One-dimensional split-ring resonator arrays. Top: All SRRs lying on a common plane ($\lambda<0$). Bottom:   All SRRs parallel but centered around a common axis ($\lambda>0$).  
  }
  \label{fig0}
\end{figure}
The configuration shown in Fig.1 with alternating positions of the slits, serves to decrease electrical dipole-dipole effects. More realistic models
consider this and other effects, but the complication added makes them intractable analytically. We will stick to model (\ref{eq1}) that, while keeping the main physics, still allows us to obtain analytical results.

On the other hand, the subject of fractional derivatives has gained increased attention in the last years.  It all started with the observation that the usual, integer-order derivative can be extended to a fractional-order derivative, that is, $(d^n/d x^n)\rightarrow (d^s/d x^s)$, for real $s$. The sth order derivative of a function $f(x)$ can be formally expressed as \cite{fractional1, fractional2, fractional3}
\be
\left({d^{s}\over{d x^{s}}}\right) f(x) = {1\over{\Gamma(1-s)}} {d\over{d x}} \int_{0}^{x} {f(x')\over{(x-x')^{s}}} dx'\label{eq:2}
\ee
where $0<s<1$. For the one-dimensional Laplacian operator $\Delta=d^2/d x^2$, its fractional form $(-\Delta)^s$ can be expressed as\cite{landkof}
\be
(-\Delta)^s U(x) = L_{s} \int {U(x)-U(y)\over{|x-y|^{n+2 s}}} dy\label{eq:3}
\ee
where,
\[
L_{s}={4^s \Gamma(s+(1/2))\over{\sqrt{\pi}|\Gamma(-s)|}},\label{eq:4}
\]
$\Gamma(x)$ is the gamma function and $0<s<1$ is called the fractional order of the Laplacian. This fractional Laplacian operator has found useful applications to fluid mechanics\cite{fluid1,fluid2}, fractional kinetics and anomalous diffusion\cite{metzler,sokolov,zaslavsky}, strange kinetics\cite{shlesinger}, fractional quantum mechanics\cite{laskin,laskin2}, Levy processes in quantum mechanics\cite{levy}, plasmas\cite{plasmas}, electrical propagation in cardiac tissue\cite{cardiac} and biological invasions\cite{invasion}.

The term $\lambda (q_{n+1} + q_{n-1})$ in Eq.(\ref{eq1}) is essentially a discrete laplacian $\Delta_{n} q_{n} = q_{n}-2 q_{n}+q_{n+1}$. Thus, we can rewrite Eq.(\ref{eq1}) as
\be
{d^2\over{d t^2}}\left( q_{n} + 2 \lambda q_{n} + \lambda \Delta_{n} q_{n}  \right) + q_{n} + \chi q_{n}^3 = 0 \label{eq2}
\ee

In this paper we examine the effects of a general fractional exponent $s$ on the 
nonlinear modes, trapping, and transport of magnetic excitations of a coupled SRR array described by a discrete, fractional Laplacian. The idea is to ascertain the degree of `robustness' of the SRR phenomenology when confronted with a perturbation not of its parameters, but of its underlying mathematical description, in the form of a nontrivial mathematical change in the form of the discrete laplacian.

{\em The model}. We now proceed to replace $\Delta_{n}$ by its fractional form $(\Delta_{n})^s$ in Eq.(\ref{eq2}). The discrete fractional laplacian is given by\cite{discrete laplacian}
\be
(-\Delta_{n})^s q_{n}=\sum_{m\neq n} K^s(n-m) (q_{n}-q_{m}),\hspace{0.5cm}0<s<1 \label{delta}
\ee
where,
\be
K^{s}(m) = L_{s}\  {\Gamma(|m|-s)\over{\Gamma(|m|+1+s)}}.\label{K}
\ee
Equation(\ref{eq1}) becomes
\begin{eqnarray}
{d^2 \over{d t^2}} (\ q_{n} + 2 \lambda q_{n}+\lambda \sum_{m\neq n}K^s(m-n) (q_{m}-q_{n})\ ) + q_{n}& &\nonumber\\
 +\ \chi\ q_{n}^3 = 0.\ \ \ \ \ \ \ \ & &\label{eq7}
\end{eqnarray}

The equation of motion Eq.(\ref{eq7}) can be derived from the dimensionless Hamiltonian 
\begin{widetext}
\begin{eqnarray}
H = \sum_{n} \left({1\over{2}}\right)\Big( q_{n}^2 + (1 + 2 \lambda) \dot{q}_{n}^2 + \lambda \dot{q}_{n} \sum_{m} K^{s} (m-n)(\dot{q}_{m}-\dot{q}_{n})\Big) + (\chi/4) q_{n}^4
\label{eq8}
\end{eqnarray}
\end{widetext}
and $\dot{q}_{n}=\partial H/\partial p_{n}$, $\dot{p}_{n}=-\partial H/\partial q_{n}$, and the definition $p_{n}=\partial H/\partial \dot{q}_{n}$. This implies that $H$ is a constant of motion: $d H/d t$= 0. Since in all of our dynamical computations later on, we will choose $q_{n}(0)=\delta_{n,0}, \dot{q}_{n}(0) = 0$, this implies $H=(1/2)+(\chi/4)$.

Now we look for stationary modes $q_{n}(t) = q_{n} \cos(\Omega t + \phi)$. After replacing this form into Eq.(\ref{eq7}) and after using Eq.(\ref{delta}), one obtains the following system of nonlinear difference equations for $\{q_{n} \}$:
\begin{eqnarray}
-\Omega^2 (\ q_{n} + 2 \lambda q_{n}+\lambda \sum_{m\neq n}K^s(m-n) (q_{m}-q_{n})\ ) + q_{n}& &\nonumber\\
 +(3/4)\ \chi\ q_{n}^3 = 0\ \ \ \ \ \ \ \ & &\label{stationary}
\end{eqnarray}
where we have also made use of the rotating wave approximation (RWA): $\cos(x)^3 \approx (3/4) \cos(x)$. This is necessary 
\begin{figure}[t]
 \includegraphics[scale=0.8]{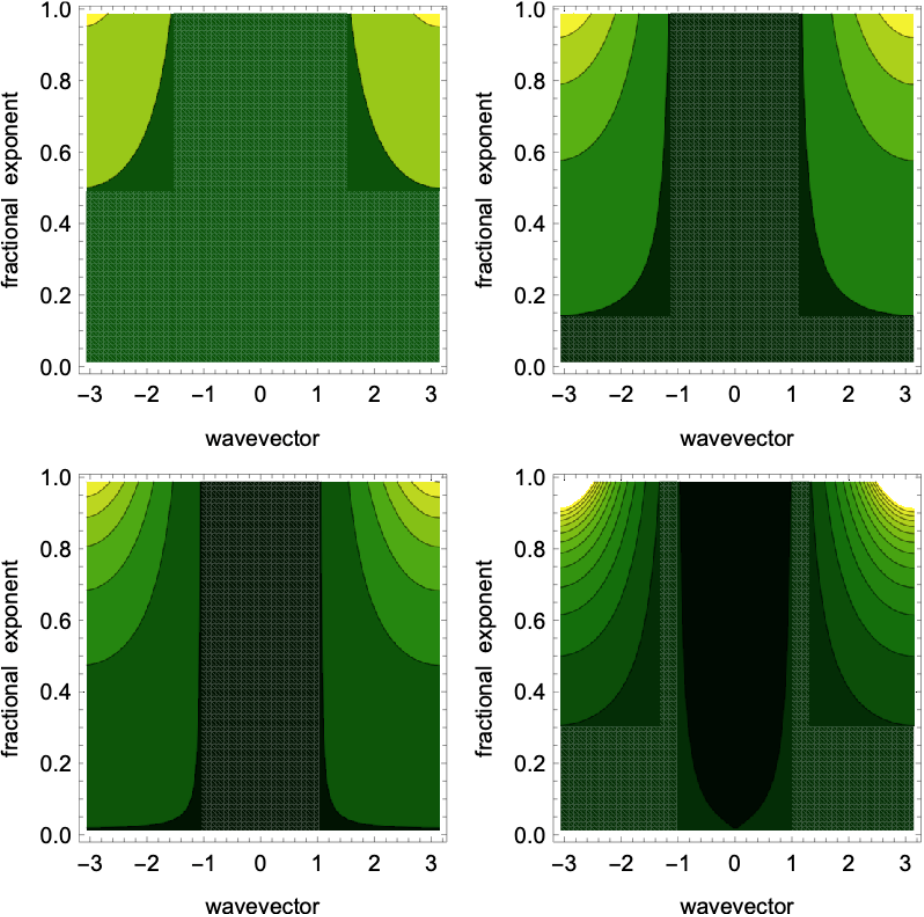}
  \caption{Left: Dispersion relation $\Omega(k,s)^2$ for several different coupling values. Top left: $\lambda=0.1$, Top right: $\lambda=0.2$, Bottom left: $\lambda=0.3$, Bottom right: $\lambda=0.5$.}
  \label{fig1}
\end{figure}
in order to obtain closed-form expressions later on. Numerically, the RWA is accurate in many cases.

Let us first look for linear waves. We set $\chi=0$ and pose a solution of the form $q_{n} = A \cos(k n)$. After simple algebra, one obtains the dispersion relation for the magnetoinductive waves, in closed form:
\be
\Omega^2 = (\  1 + 2 \lambda - 4 \lambda \sum_{m=1}^{\infty} K^s(m)\ \sin^2[(1/2) k\ m]) \ )^{-1}\label{dispersion}
\ee
The condition $\Omega^2 >0$ leads to a constraint on the coupling values:
\be
\lambda >  \left(\ 4\sum_{m=1}^{\infty}K^s(m)\ \sin^2[(1/2) k\ m]-2\ \right)^{-1}.
\label{lambdac}
\ee

Since the extremes of the band occurs at $k=0,\pi$, we evaluate Eq.(\ref{lambdac}) at these points, obtaining the critical coupling conditions   as functions of $s$:
\be
\hspace{1.0 cm}\lambda > -1/2 \ \ \mbox{and}\ \ \lambda <\left( 4 \sum_{m\ odd}^{\infty} K^{s}(m) - 2 \right)^{-1}
\label{lambda_critico}
\ee 
In particular, for $s\rightarrow 1$ one recovers the well-known condition $|\lambda|<1/2$. A further reduction on Eqs.(\ref{dispersion}) and (\ref{lambda_critico}) is possible, in terms of well-known special functions:
\begin{widetext}
\begin{eqnarray}
\Omega^2 &=& \nonumber\\
        &  &\Big(\ 1 + \lambda \Big(\ 2 + { 2 e^{-i k}\Gamma(1-s)\Gamma(2 s)(-e^{i k}(1+s) + s\  _{2}F_{1}(1,1-s,2+s,e^{-i k}) + e^{2 i k} s\ _2F_1 (1,1-s,2+s,e^{i k})) )\over{|\Gamma(-s)| \Gamma(1+s) \Gamma(2+s)}}\ \Big) \Big)^{-1}\ \ \ \ \ \label{om2}
\end{eqnarray}
\end{widetext}
where $\Gamma(x)$ is the gamma function and $_2F_1(a,b,c;z)$ is the hypergeometric function. The critical coupling values can be reduced to
\be
\lambda > -1/2\ \ \ \ \mbox{and}\ \ \ \ \lambda < -{1\over{2 + 4 ^{s}\ \mbox{sign}\ (\Gamma(-s))}}
\ee
Examination of this expression reveals that, if we choose to stick to $|\lambda|<1/2$ the dispersion will always be real for all values $0<s<1$. In that case, we have well-defined `fractional magnetoinductive waves'.
Figure 2 shows the dispersion relation $\Omega^2$ as a function of $(k,s)$ for several coupling values $\lambda$. The bandwidth increases with $s$, while the average gradient increases with increasing $\lambda$. Figure 3 shows density plots for the mode profiles, for several values of the fractional exponent s. Here, for a given s, we stack the mode profiles one after the other according to their eigenvalues. As can be seen, an increase in the fractional exponent, increases the bandwidth. Figure \ref{critical} shows the allowed regions in coupling-exponent space,  for propagation of these fractional magnetoinductive waves.
\begin{figure}[t]
 \includegraphics[scale=0.35]{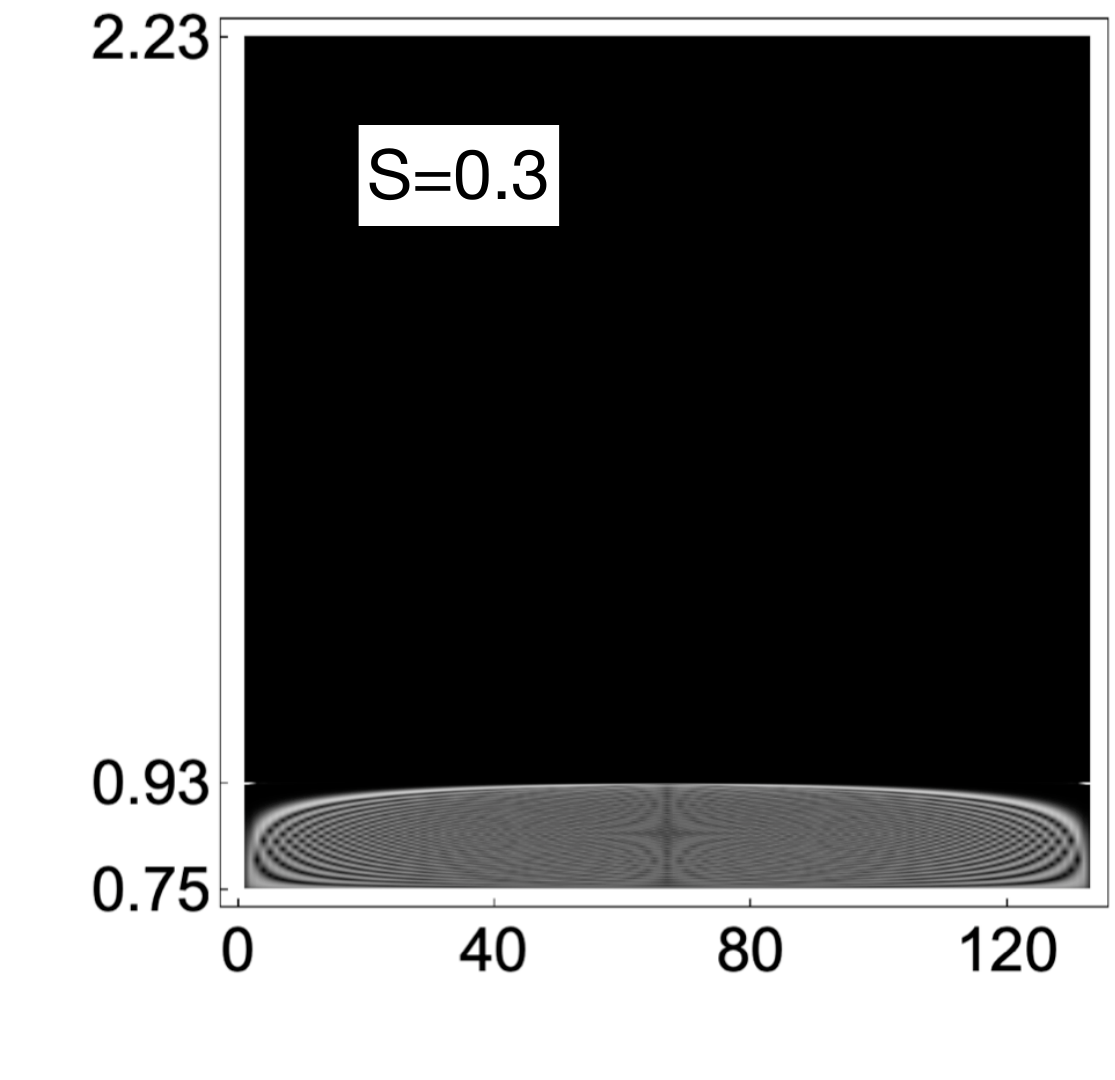}
 \includegraphics[scale=0.35]{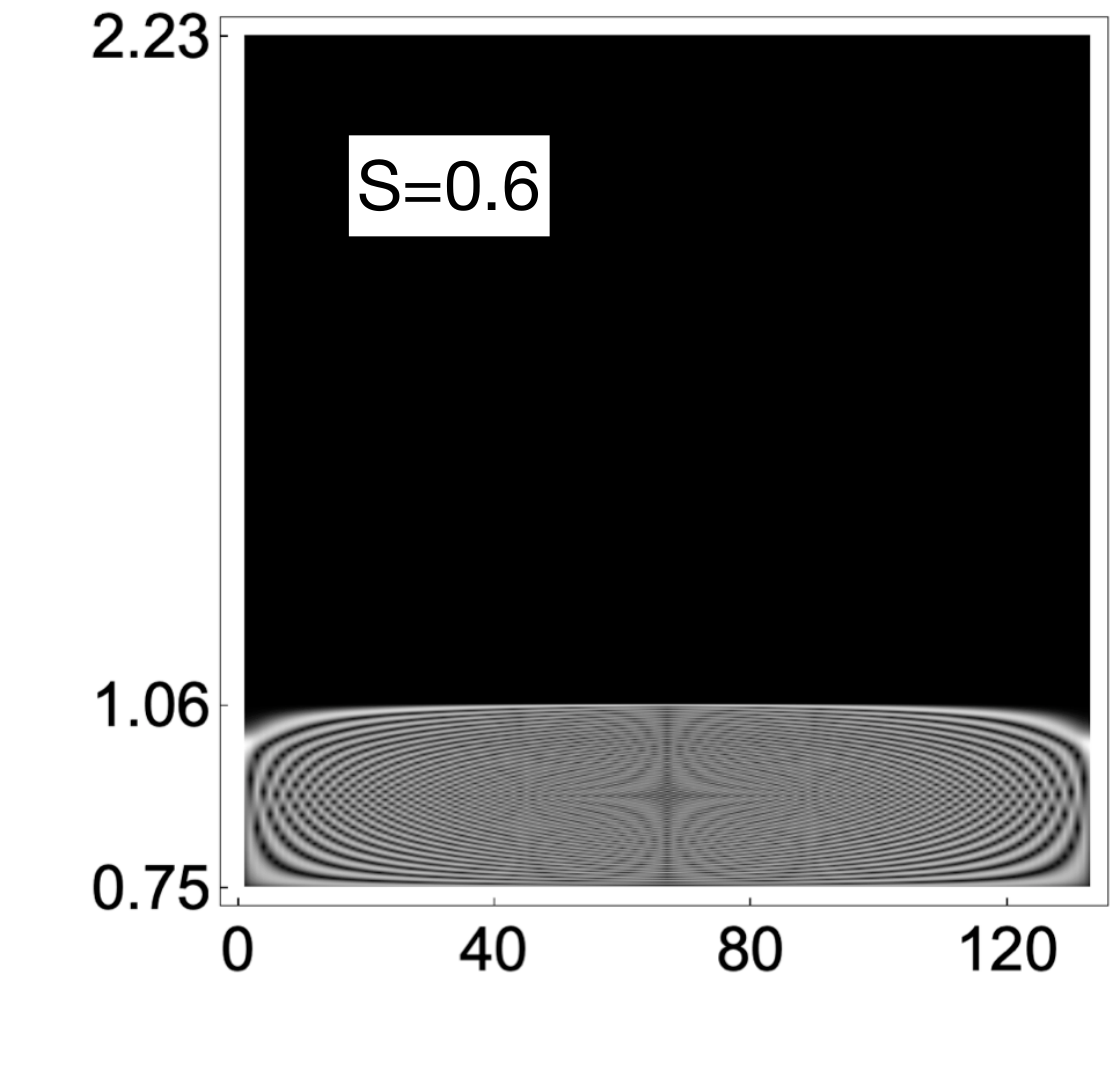}\\
 \includegraphics[scale=0.35]{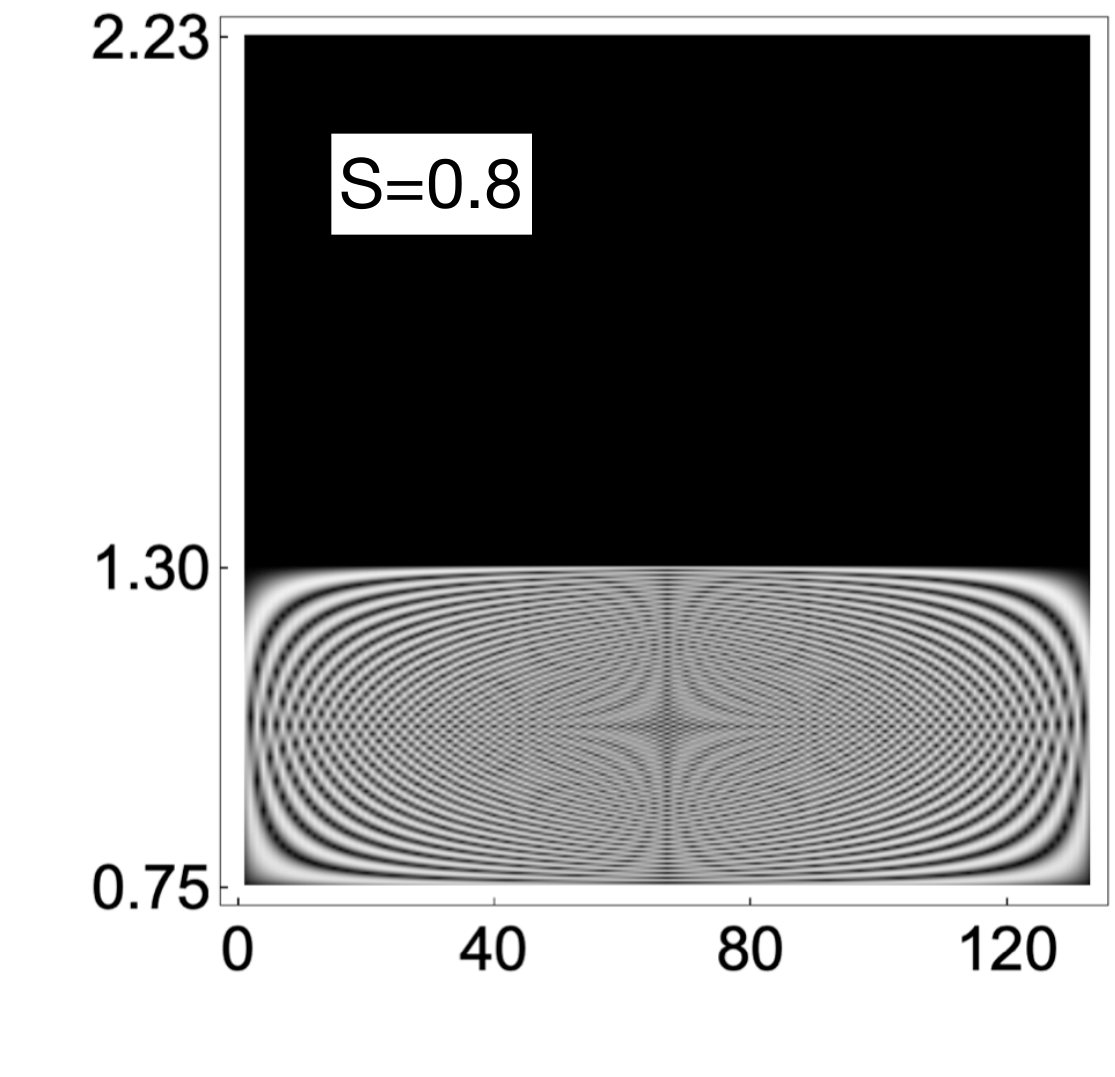}
 \includegraphics[scale=0.35]{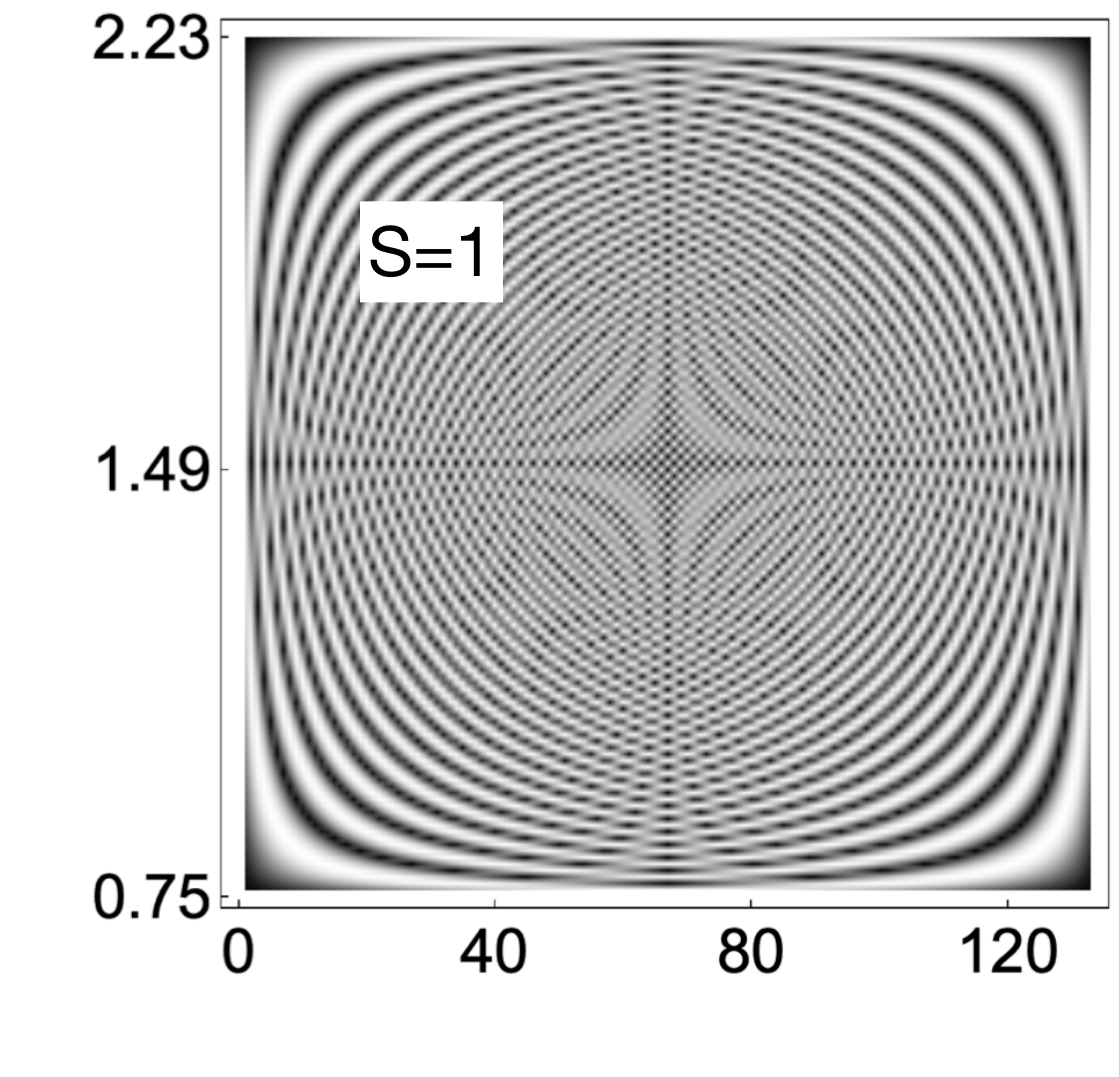}
  \caption{Density plot of the spatial profiles $q_{n}^2$ of the linear modes ordered according to their eigenvalue. Horizontal axis denota the sites of the array, while the vertical axis contains the frequencies in increasing order ($N = 133, \lambda=0.4$).}
  \label{fig2}
\end{figure}

{\em Root mean square (RMS) displacement}. A common way to quantify the degree of mobility of an excitation, is by means of its mean square displacement $\langle n^2 \rangle$. In our case,$\langle n^2 \rangle$ measure the spread of the charge among the rings:
\be
\langle n^2 \rangle = \sum_{n} n^2 |q_{n}(t)|^2 / \sum_{n} |q_{n}(t)|^2\label{n2}
\ee
For a completely localized initial charge $q_{n}(0)=A\ \delta_{n 0}$ and no currents, $(d q_{n}/d t)(0)=0$, we have formally
\be q_{n}(t) = (A/4 \pi) \int_{-\pi}^{\pi} e^{i (k n-\Omega_{k})t} dk + (A/4 \pi) \int_{-\pi}^{\pi} e^{i (k n+\Omega_{k})t}
\ee
where $\Omega_{k}$ is given by Eq.(\ref{om2}). After replacing this form for $q_{n}(t)$ into Eq.(\ref{n2}), one obtains after some algebra, a closed form expression for $\langle n^2 \rangle$:
\be
\langle n^2 \rangle = {(1/2 \pi)\int_{-\pi}^{\pi}d k (d \Omega_{k}/d k)^2 (1 - \cos(2\ \Omega_{k}\ t))\ t^2 
\over{1 + (1/2\pi) \int_{-\pi}^{\pi} d k\ \cos(2\ \Omega_{k}\ t)}}\label{n2closed}
\ee
As we can see from the structure of Eq.(\ref{n2closed}), as time $t$ increases, the contributions from the cosine terms to the integrals decrease and, at long times, $\langle n^2\rangle$ approaches a ballistic behavior
\be
\langle n^2 \rangle \rightarrow \left[ {1\over{2 \pi}} \int_{-\pi}^{\pi} \left( {d \Omega(k)\over{d k}}\right)^2\ dk\right]\ t^2,\label{RMS}
\ee
Since the transport exponent is defined as the one corresponding to the dominant behavior at long times, we can say that the transport in our system is ballistic: $\langle n^2 \rangle \sim g(s) t^2$, where we can identify $\sqrt{g(s)}$ as a kind of characteristic `speed' for the ballistic propagation.

Results are shown in Fig.\ref{critical}, which shows $g(s)$ for several coupling values. We see that a well-defined speed is only possible for exponents greater that a minimum one, $s> 0.253$. We also observe for a given $s$, this speed increases with the coupling $\lambda$, which is not surprising since a larger coupling facilitates the motion between neighboring sites. For very small $\lambda$, the speed seems to converge to a set of closely-spaced values. Also, for $\lambda < 1/2$, $g(s)$ remains finite throughout the entire fractional domain. However, for $\lambda>1/2$,  
$g(s)$ diverges at some $s$ value.   
\begin{figure}[b]
 \includegraphics[scale=0.45]{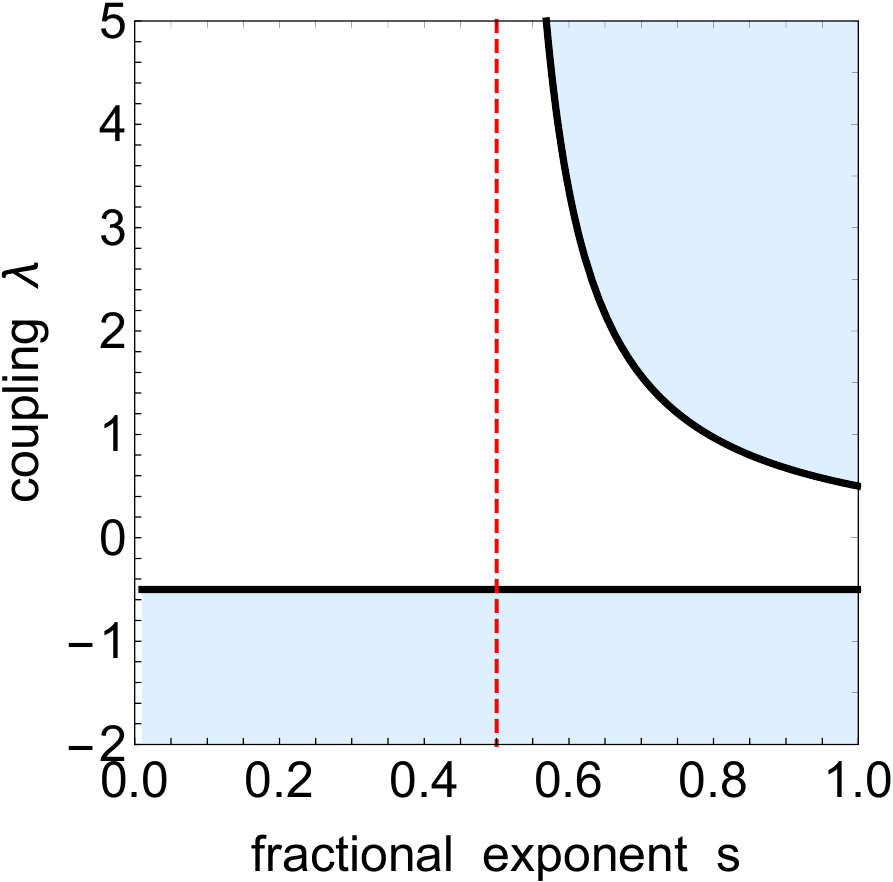}
 \includegraphics[scale=0.195]{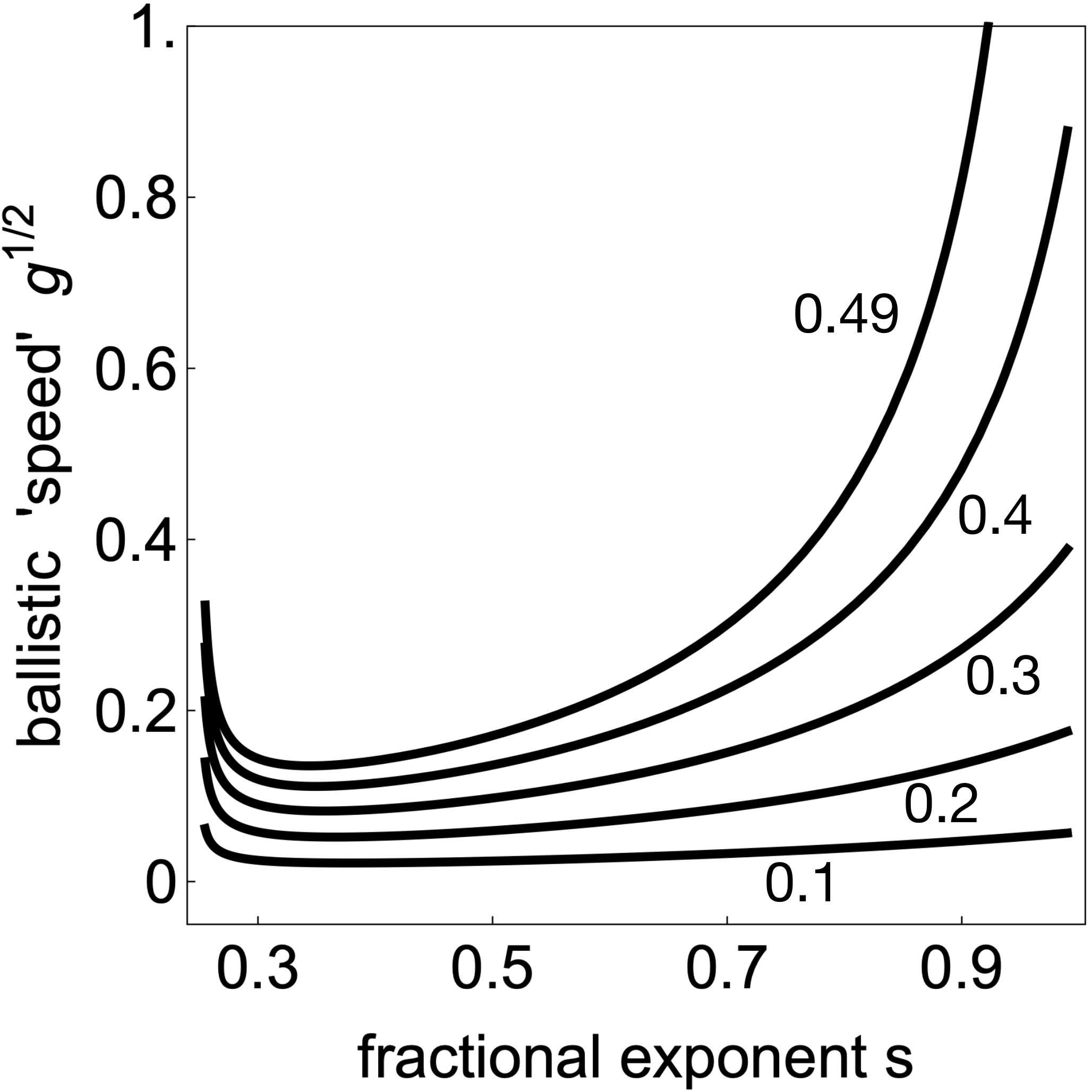}
  \caption{Left: Critical coupling as a function of the fractional exponent. Inside the shaded region, no wave propagation is allowed. 
  Right: Ballistic speed $g(s)^{1/2}$ as a function of the fractional exponent $s$, for several coupling values $\lambda$. As $\lambda\rightarrow 1/2$, $g(s)$ diverges at $s\rightarrow 1$.}
  \label{critical}
\end{figure}

\begin{figure}[t]
 \includegraphics[scale=0.5]{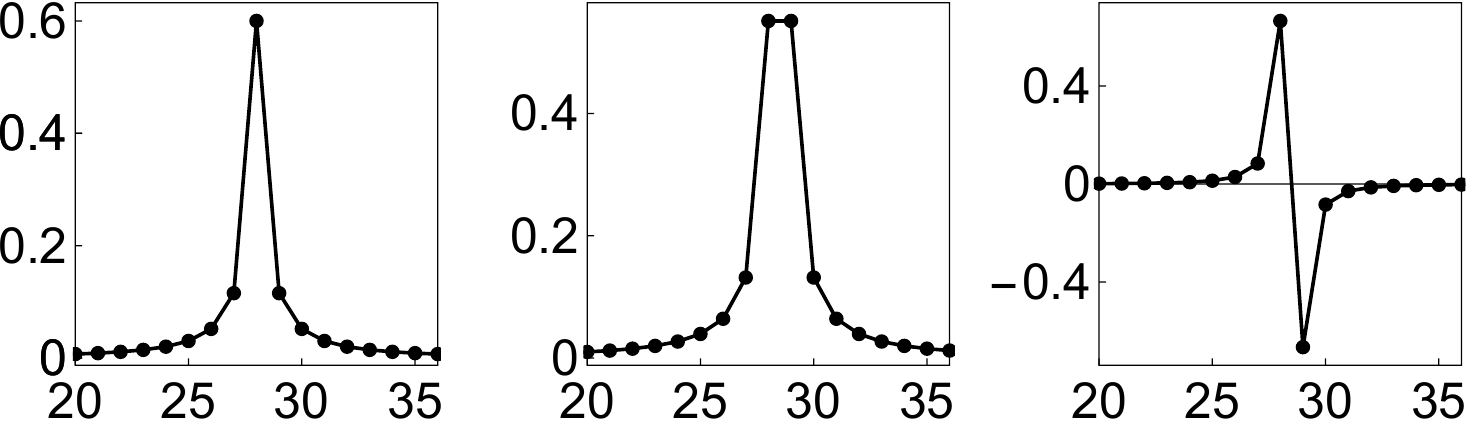}\\
 \includegraphics[scale=0.5]{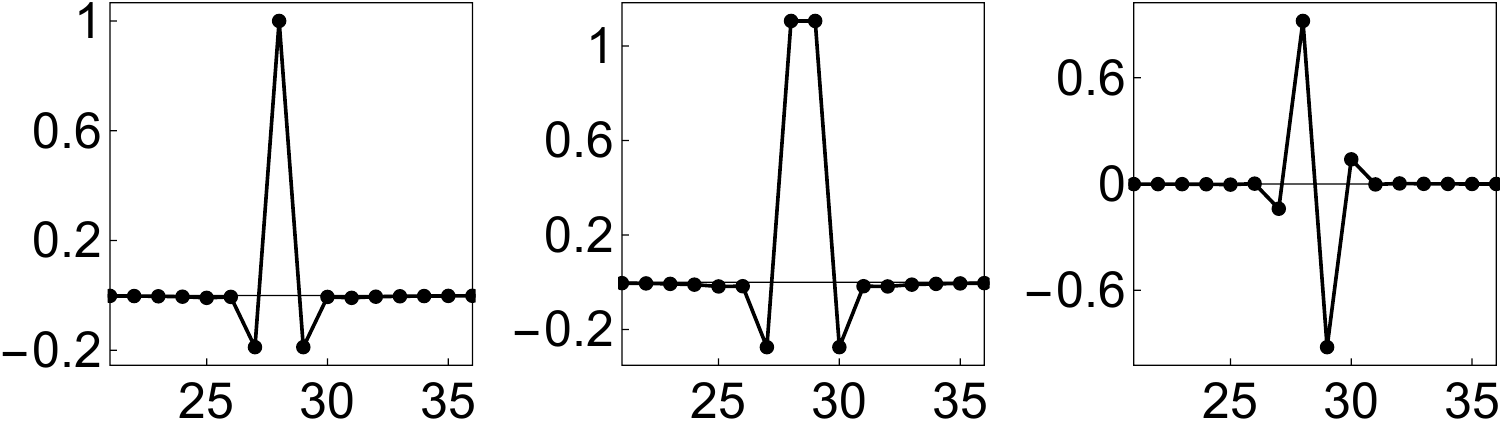}\\
  \includegraphics[scale=0.5]{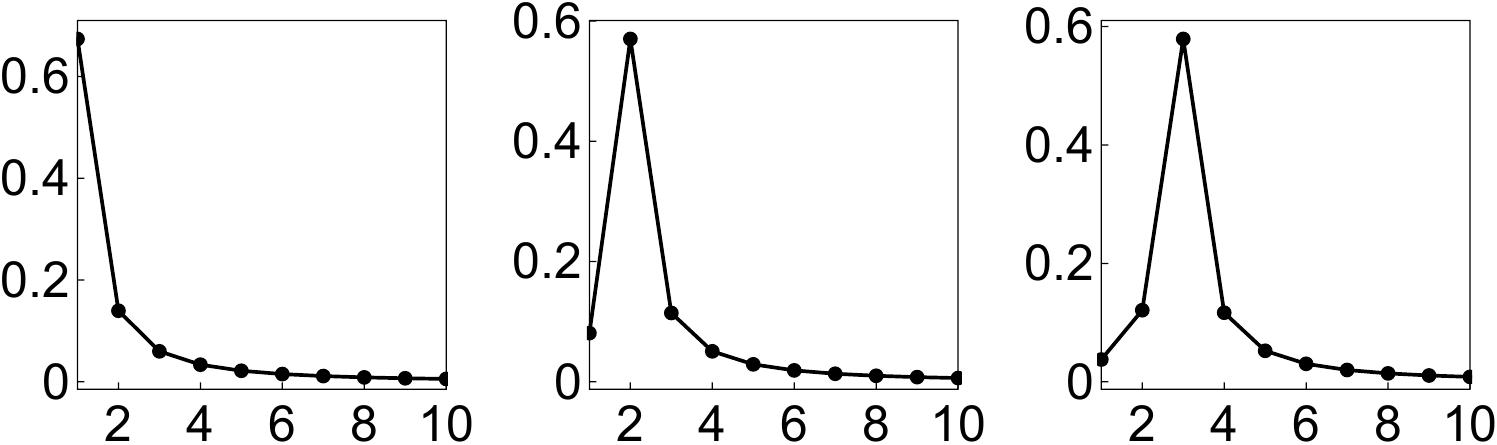}\\
   \includegraphics[scale=0.5]{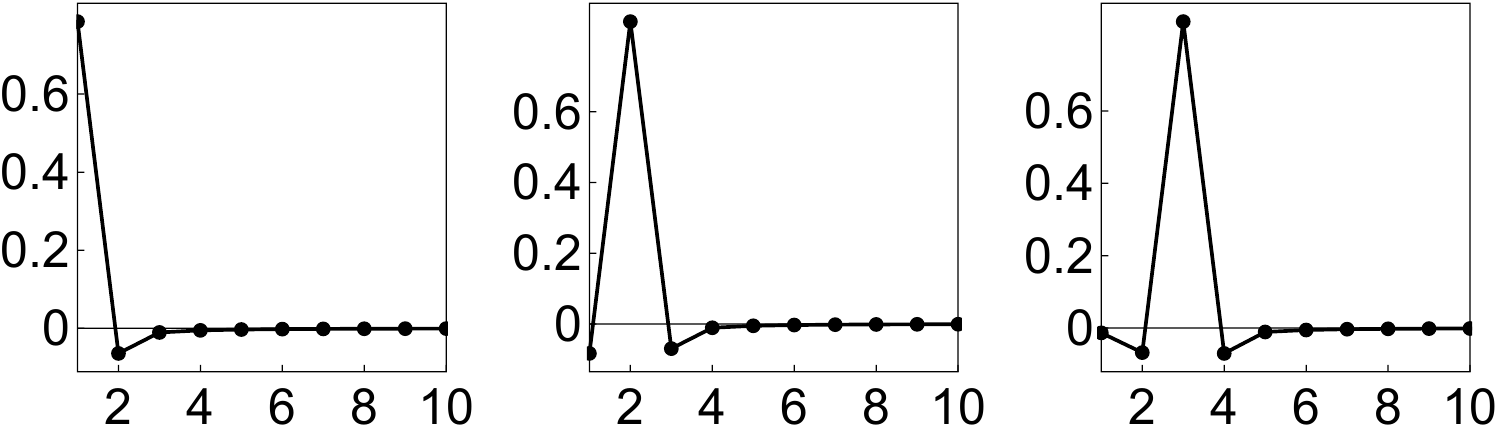}
 
  \caption{Examples of low-lying nonlinear modes of the SRR array for fractional exponent $s=0.4$. First row: Bulk modes for $\lambda=0.4$ and $\chi=-1$. From left to right: `odd' mode, even' mode and `twisted' modes, respectively. Second row: Bulk modes for $\lambda=0.4$ and $\chi=1$. Third row: Modes at and below the surface for $\lambda=0.4$ and $\chi=1$. Fourth row: Modes at and below the surface for $\lambda=0.4$ and $\chi=-1$. A reduced portion of the entire lattice is shown ($N=55$).}
  \label{NLmodes}
\end{figure}

{\em Nonlinear modes}. We now look at stationary modes when $\chi\neq 0$. Equations (\ref{stationary}) constitute a system of nonlinear coupled difference equations, with a nonlocal coupling. The form of the cubic terms originates from the insertion of a nonlinear (Kerr) dielectric inside the slit of the capacitors, in the weak nonlinear limit. Numerical solutions are obtained by the use of a multidimensional Newton-Raphson scheme, using as a seed the form obtained from the decoupled limit $(\lambda\rightarrow 0)$, also known as the anticontinuous limit. We employ open boundary conditions and examine two-mode families, ``bulk'' modes, which are located far from the boundaries, and ``surface modes'' when the mode is located at or near one of the ends of the array. Figure \ref{NLmodes}
\begin{figure}[t]
 \includegraphics[scale=0.24]{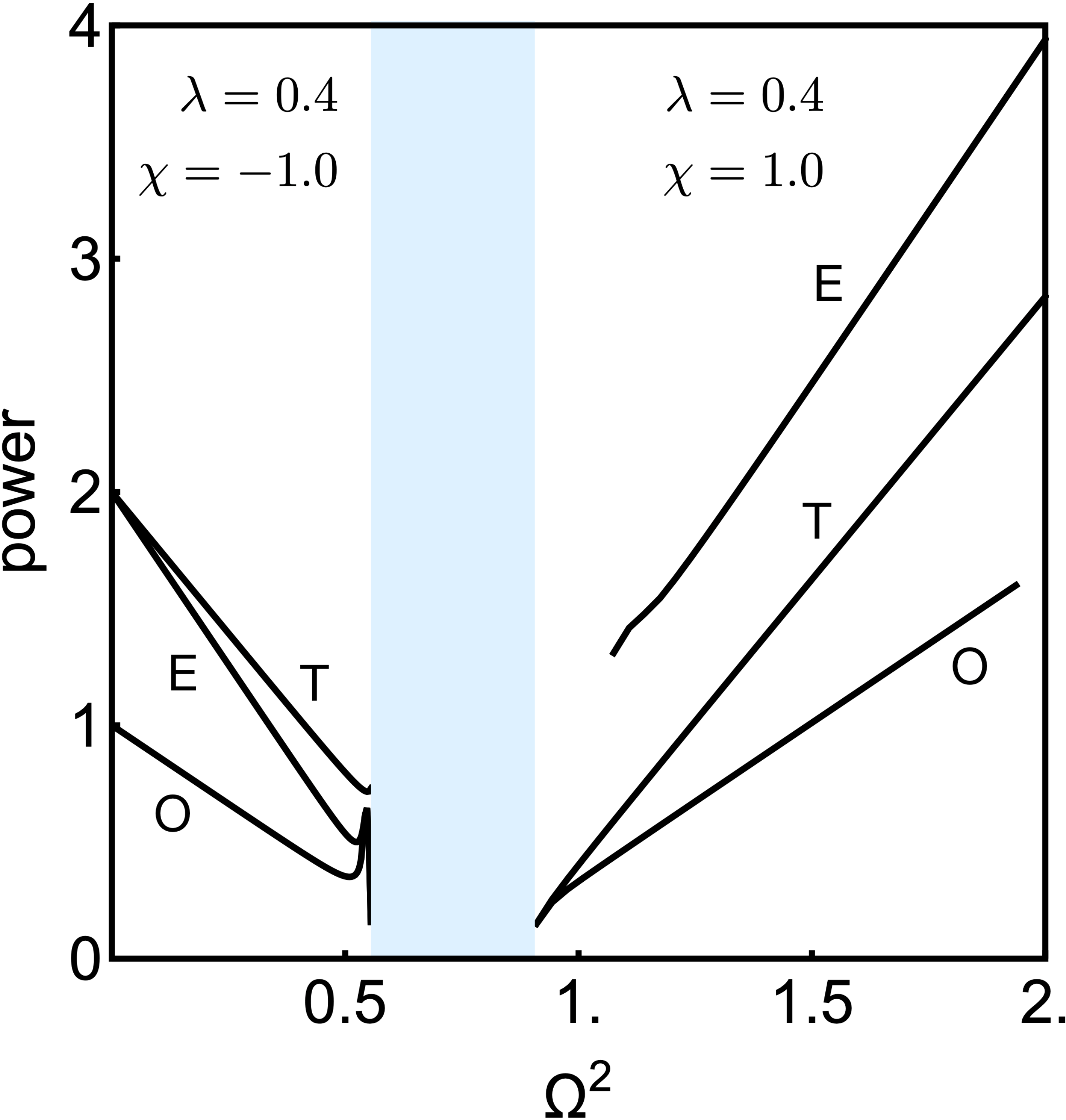}
 \hspace{0.15cm}
 \includegraphics[scale=0.24]{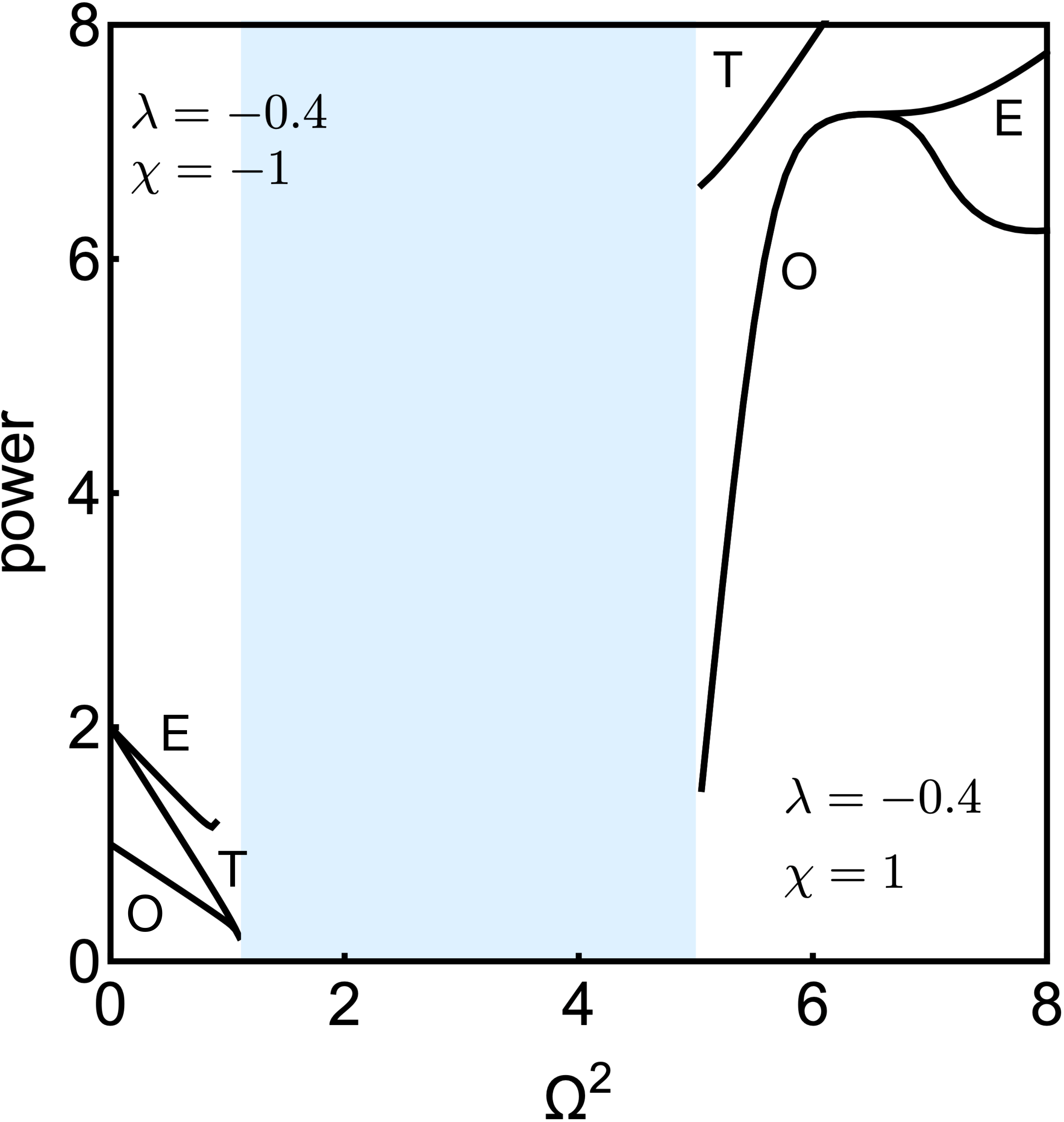}
  \caption{Bifurcation diagram for the power content versus mode frequency for some families of nonlinear modes, for fractional exponent $s=0.4$. Curves marked 'O',`E' and `T' denote odd, even and twisted modes, respectively. The shaded regions mark the position of the corresponding linear bands.}
  \label{bifurcations}
\end{figure}
shows some low lying nonlinear bulk modes for $s=0.4$, for the two possible signs of the magnetic coupling $\lambda$. Figure \ref{bifurcations} shows bifurcation diagrams of the mode families shown in Fig. \ref{NLmodes}. 
These are obtained, for instance,  by plotting the sum of the electric power content of the uncoupled rings,  $\sum_{n} q_{n}^2$, as a function of the mode frequency. We observe that the fundamental mode reaches down to the band edge, at low powers,  while the rest of the modes lie outside the band for any amount of power. The behavior evidenced in Figs. \ref{NLmodes}, \ref{bifurcations} is reminiscent of the behavior seen in the discrete nonlinear Schr\"{o}dinger equation\cite{fractional_dnls}.  For surface modes (not shown), it is possible to find a family of fundamental surface solutions located at ever-increasing distances from the boundary. After a few sites from the boundary, they become indistinguishable from  bulk modes. Also, it should be mentioned that all these surface modes need a minimum value of power strength to exist. This can be explained from the observation that when we have a mode at the surface, the concentration of magnetic energy produces an effective surface impurity at the edge. The rest of the lattice is left with a very small magnetic energy, i.e., it becomes an effective linear lattice. As is well-known, a linear lattice with an impurity at the edge needs a minimum amount of `strength' to give rise to a localized mode. By the same token, in the case of a bulk mode, the nonlinearity is concentrated at the center, producing an effective impurity embedded in an effective linear lattice. In this case, a localized mode is possible for any impurity strength. This is seen in the bifurcation curve for the fundamental mode (Fig.\ref{bifurcations}).

The existence of these localized nonlinear modes implies a spatial concentration of charge and current, which in turn produces a spatial localization of magnetic energy. This phenomenology of mode localization seems `robust' against the particular value of a fractional exponent.

\begin{figure}[t]
 \includegraphics[scale=0.8]{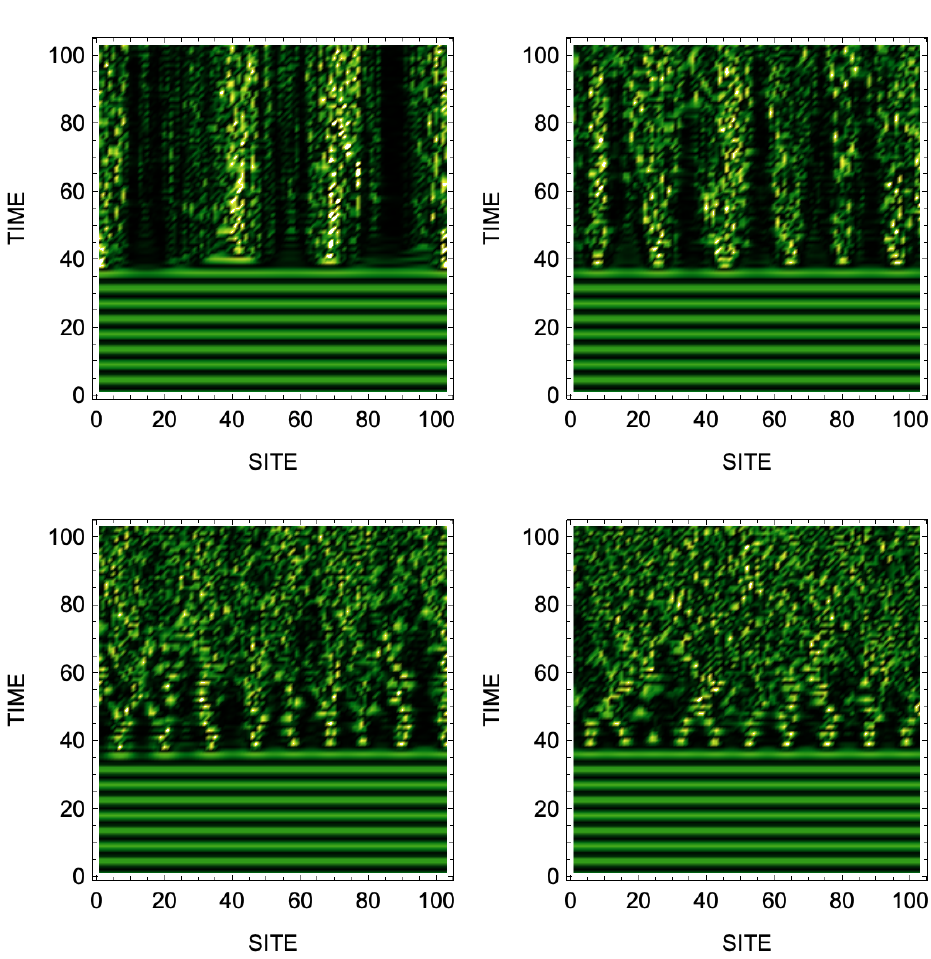}
  \caption{Modulational instability of a given uniform unstable profile for several fractional exponents. Top left: $s=0.2$, Top right: $s=0.5$, Bottom left: $s=0.9$, Bottom right: $s=1$. ($N=103$, evolution time=$50$, $\chi=1$, amplitude=$4$, $\lambda=-0.4$)}
  \label{seven}
\end{figure}

{\em Modulational Stability}. As we have seen, our system is able to support the existence of localized magnetic energy excitations, which can exist around any ring in principle. We wonder now about the possible existence and stability of nonlinear extended uniform excitations. To do that, we consider Eq.   
(\ref{eq2}) in the form:
\begin{eqnarray}
&   &{d^2\over{d t^2}}(\ q_{n} + 2 \lambda q_{n}+\lambda \sum_{m\neq n}K^s(m-n) (q_{m}-q_{n})\ ) + q_{n}\nonumber\\
&   & +\ \chi\ q_{n}^3 =  0 \label{modulational}
\end{eqnarray}
Now we pose a solution in the form of a uniform profile $q_{n}(t) = A \cos(\Omega t + \phi)$. Inserting this solution form into Eq.(\ref{modulational}), followed by the RWA approximation, we obtain
\be
\Omega^2 = {1 + (3/2) \chi A^2\over{1 + 2 \lambda}}.
\ee
Thus, the uniform profile is possible only if $1+(3/2) \chi\ |A|^2 > 0$ (we are assuming $|\lambda|<1/2$). We compute $q_{n}(t)^2$ numerically from $t=0$ up to a $t=t_{max}$ and examine its numerical stability. Figure \ref{seven} shows some examples of results obtained from this procedure, where we plot the time-evolution of the SRRs spatial profile, for several fractional exponents. 
From Fig.\ref{seven} we see that the evolution of a uniform unstable profile consists in the persistence of the initial profile up to some time, where it decays into radiation, after a very short transient. The most interesting feature is the decay of the unstable state into filamentary structures that, as time runs, merge and ultimately form undifferentiated radiation. This contrast with other (no fractional) models\cite{models},  where these filaments are also observed that persist in time and give rise to discrete solitons embedded in radiation. This fact has been proposed as a mechanism for producing discrete solitons empirically. 
\begin{figure}[t]
 \includegraphics[scale=0.55]{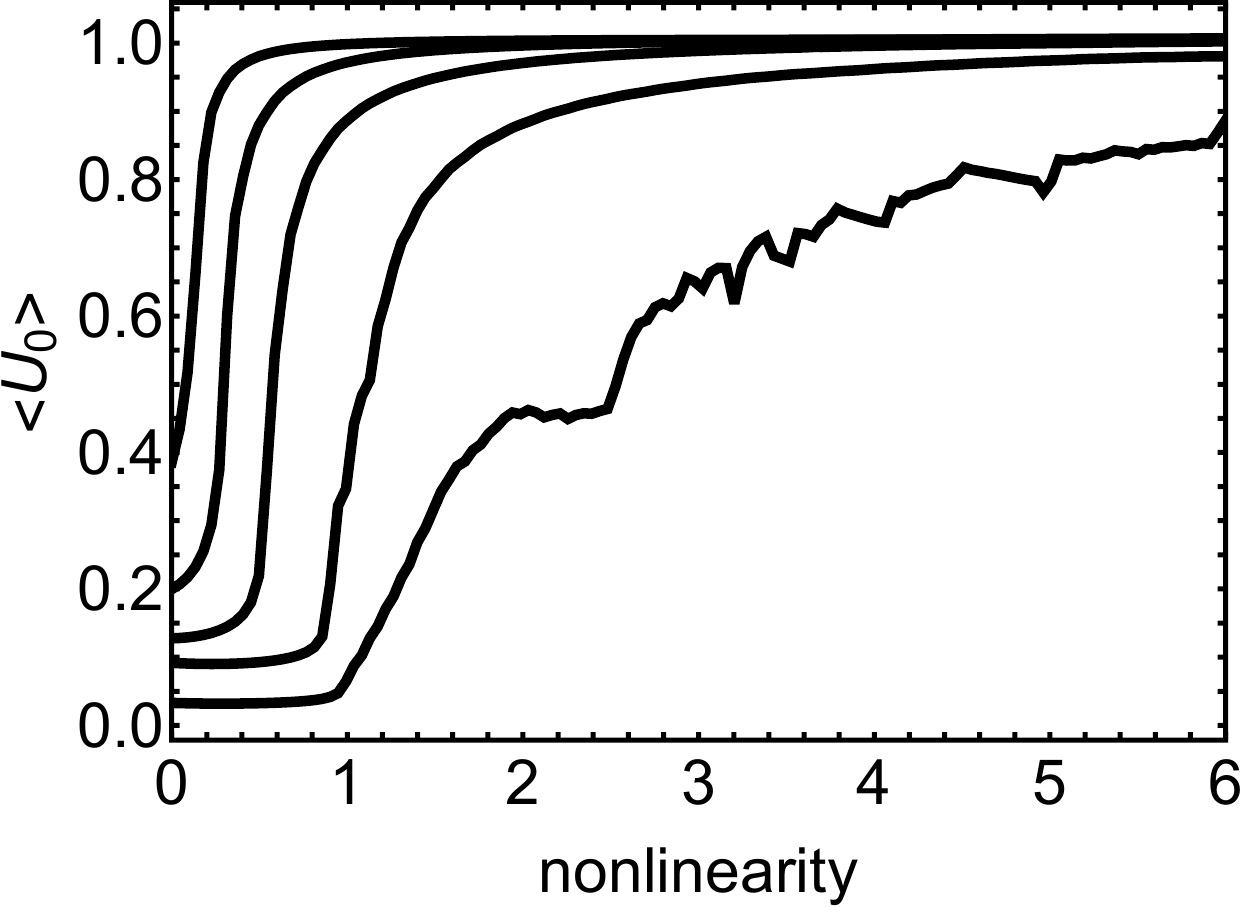}
  \caption{Long-time average probability of the magnetic energy remaining at  the initial ring, versus nonlinearity strength, for different fractional exponents. From left to right $s=0.2, 0.4, 0.6, 0.8$ and $1.0$.}
  \label{eight}
\end{figure}

{\em Dynamical selftrapping}. We now examine the possible presence of magnetic energy selftrapping of an initially localized excitation. The local nonlinearity favors an accumulation of all energy at a given position. This in turn, causes an effective impurity embedded in an effective linear lattice. As is well known from linear theory, this creates a localized impurity mode which in our case would correspond to a selftrapped state.

We create an initial magnetic localized excitation by inducing a single current around one of the rings (at $n=n_{0}$), by means of an adequate antenna. We are interested in monitoring the amount of magnetic energy that remains at long times at the initial ring. To quantify the trapping of magnetic energy, we use the long-time average of the fraction of magnetic energy residing at the initial site:
\be 
\langle U_{0} \rangle = {1\over{T}} \int_{0}^{T} (h_{0}/H)\ dt,
\ee
where
\begin{eqnarray}
h_{0}&=&(1/2) ( q_{0}^2+(1+2 \lambda) \dot{q}_{0}^2 +\nonumber\\
     & & + \lambda \dot{q}_{0}\sum_{m} K(m) (\dot{q}_{m}-\dot{q}_{0}))+(\chi/4) q_{0}^4,
\end{eqnarray}
and $H=(1/2)+(\chi/4)$,  as shown before. Figure \ref{eight} shows $\langle U_{0} \rangle$ as a function of $\chi$, for several fractional exponents.
For a given value of $s$, $\langle U_{0} \rangle$ increases monotonically with $\chi$, with a more or less well-defined transition around some $\chi$-value. We also see that all selftrapping curves lie one on top of the other, with the highest one corresponding to the smallest $s$ value.  In the linear limit ($\chi=0$), we observe a non-zero value for $\langle U_{0} \rangle$. This type of `linear' trapping is stronger as $s$ decreases.

{\em Conclusions}. We have examined the physics of an extended model of a coupled split-ring resonator array, by using a fractional form for the Laplacian operator. A closed-form expression for the dispersion of plane waves as a function of the fractional exponent is found, in terms of well-known special functions. The critical coupling among rings beyond which no magnetoinductive waves can exist was also obtained in closed form. The diffusion of an initially localized excitation was shown analytically to be ballistic at long time.
This result is shown to be generic for a whole family of tight-binding models obeying very general conditions. The selftrapping of an initial excitation shows the same features found in other nonlinear models, except for the presence of finite trapping in the linear limit for a finite fractional exponent.  This can be understood as the consequence of the long-range coupling that is present when $s\rightarrow 0$. In this limit, the bandwidth converges to zero, and all sites are coupled to each other with similar couplings. The near-degenerate spectrum that results,  leads to partial trapping of an initially localized excitation, as shown before in condensed matter for the simplex model\cite{kenkre}.

Interestingly, the nonlinear phenomenology found in this model is similar for all fractional exponents, and also similar to the phenomenology found in other non-fractional models, as the DNLS.  In other words, the existence and stability of nonlinear modes would be more or less generic to systems with or without fractionality. This is understandable since a fractional exponent merely gives rise to a nonlocal coupling among rings. Thus, we conjecture that in 2D arrays similar results will be obtained. Results on this will be reported elsewhere. 

\acknowledgments
This work was supported by Fondecyt Grant 1160177.

\end{document}